




\message{<< Assuming 8.5" x 11" paper >>}    

\magnification=\magstep1	          
\raggedbottom

\parskip=9pt

\def\singlespace{\baselineskip=12pt}      
\def\sesquispace{\baselineskip=16pt}      


 \let\miguu=\footnote
 \def\footnote#1#2{{$\,$\parindent=9pt\baselineskip=13pt%
 \miguu{#1}{#2\vskip -7truept}}}

\def\alfa{\alpha}
\font\openface=msbm10 at10pt
\def\Minkowski     {{\hbox{\openface M}}}
\def\NaturalNumbers{{\hbox{\openface N}}}

\def\tilde{\widetilde}		

\font\titlefont=cmb10 scaled\magstep2 
\font\subheadfont=cmssi10 scaled\magstep1 
\def\author#1 {\medskip\centerline{\it #1}\bigskip}
\def\address#1{\centerline{\it #1}\smallskip}
\def\furtheraddress#1{\centerline{\it and}\smallskip\centerline{\it #1}\smallskip}
\def\email#1{\smallskip\centerline{\it address for email: #1}}
\def\eprint#1{{\tt #1}}
\def\reference{\hangindent=1pc\hangafter=1} 
\def\ref{\reference}
\def\AbstractBegins
{
 \singlespace                                        
 \bigskip\leftskip=1.5truecm\rightskip=1.5truecm     
 \centerline{\bf Abstract}
 \smallskip
 \noindent	
 } 
\def\AbstractEnds
{
 \bigskip\leftskip=0truecm\rightskip=0truecm       
 }

\def\ReferencesBegin
{
 \singlespace					   
 \vskip 0.5truein
 \centerline           {\bf References}
 \par\nobreak
 \medskip
 \noindent
 \parindent=2pt
 \parskip=6pt			
 }

\def\journaldata#1#2#3#4{{\it #1\/}\phantom{--}{\bf #2$\,$:} $\!$#3 (#4)}

\def\sepref{\baselineskip=18pt \hfil\break \baselineskip=12pt}

\def\subsection #1 {\medskip\noindent{\subheadfont #1
 }\par\nobreak\noindent}

\def\linebreak{\hfil\break}
\def\lbr{\linebreak}

\def\dash{ --- }



\phantom{}




\sesquispace

\centerline {\titlefont Relativity theory does not imply that the future } 
\centerline{{\titlefont already exists: a counterexample}\footnote{$^{^{\displaystyle\star}}$}%
{ to appear in 
 Vesselin Petkov (editor),
 {\it Relativity and the Dimensionality of the World},
 a book in the series ``Fundamental Theories of Physics''
 (Springer 2007, in press)}}

\bigskip


\singlespace			        

\author{Rafael D. Sorkin}

\address{Perimeter Institute, 31 Caroline Street North, Waterloo ON, N2L 2Y5 Canada}

\furtheraddress{Department of Physics, Syracuse University, Syracuse, NY 13244-1130, U.S.A.}

\email{sorkin@physics.syr.edu}

\AbstractBegins                              
It is often said that the relativistic fusion of time with space
rules out genuine change or ``becoming''.
I offer the classical sequential growth models of causal set theory as
counterexamples.
\AbstractEnds


\sesquispace

\vskip 30pt

\noindent
1. Can one hold a ``four-dimensional'' point of view and still maintain
consistently that things really happen?  Is a spacetime perspective
compatible with the idea of ``becoming''?  Many authors have denied such
a possibility, leaving us to choose between a static conception of
reality and a return to the pre-relativistic notion of linear time.  In
contrast, I want to offer a concrete example \dash a theoretical model of
causal set dynamics \dash that illustrates the possibility of a positive
answer to the above questions, according to which reality is more
naturally seen as a ``growing being'' than as a ``static thing''.

Of course, one might doubt whether the static and dynamic conceptions of
reality differ in more than words, given that the distinction between them
does not seem to find a home in the mathematics of general relativity.
Do the Einstein equations look any different when they are viewed
``dynamically'' rather than ``under the aspect of eternity''?, 
a skeptic might ask.  
Or just because the psychological feeling of ``the now''
impresses itself on our minds, should that really matter
to us as physicists?  
Such questions threaten to lead off into  
an impassable terrain of metaphysics, metamathematics, and the
meaning of meaning.  
But 
this does not mean that the questions
``being or happening?'', ``static or dynamic?''
lack practical significance for the working
scientist, because the answers one gives will inform the {\it direction} in
which one searches for new theoretical structures.  Thus,
for example,
the dynamical scheme that I will be using for illustration 
has sprung from the search for a theory of quantum gravity.
We will see that it does provide a sort of mathematical home for the idea
of becoming (as a process of growth or birth) 
and that, conversely, 
one
 would have been hard-pressed to
arrive at such a dynamical scheme without starting from the idea of a
Markov process unfolding in time.

\noindent
2. The model I'm referring to is 
that of {\it classical sequential growth} (CSG) 
regarded as 
a ``law of motion'' or ``dynamical law'' 
for causal sets.
You can find in references [1] and [2] 
a full mathematical description of the model, 
and in [3] an account of how one resolves in that context the
complex of conceptual difficulties known to workers in quantum gravity
as ``the problem of time''.
Here I will just try to summarize the basic ideas with an emphasis on
those aspects most germane to the present discussion.

A CSG model describes a stochastic process in which elements $e$ of the
growing causal set $C$ are born one by one, each with a definite subset
of the already born elements as its ancestors.
If one records the ancestral relationships among a set of elements
produced in this manner, 
the resulting ``family tree'' will be an instance of a 
{\it causal set\/} [4].
Mathematically characterized, 
what one obtains
is,
more precisely,
a past-finite partial order
in which $x$ precedes $y$ iff $x$ is an ancestor of $y$,
$x$ and $y$ being arbitrary elements of $C$. 
In a CSG dynamics, the specific births that occur
are (with trivial exceptions) not determined in advance; 
rather they happen stochastically, 
in such a manner as to define a Markov process.  
A specific member of the family of CSG models 
is determined by the set of
transition probabilities of the Markov process; 
and these in turn
can be expressed in terms of the basic parameters or 
``coupling constants'' of the theory, 
as explained in [1].


For example, let $e_0$ be the first-born element, $e_1$ the next
born, etc.  The birth of $e_0$ can be construed as a transition from
the empty causal set to the (unique) causal set of one element, and it
occurs with probability 1.  
The next birth,
however, 
can occur in two different ways: either $e_0$ will be an ancestor of
$e_1$ (written $e_0\prec e_1$), or it will not; and each of these two
events will happen, in general, with non-zero probability.  
After the third birth the possible outcomes number five, 
and at subsequent stages 
the number of possible causal sets
rises
rapidly.  
After the fourth birth, one can have any of 16
non-isomorphic causal sets, while after the tenth there are already over
two million distinct possibilities (2567284 to be precise).  How likely
any one of these possibilities is to be realized depends on how the
parameters of the model are chosen.  
At one extreme, each new element acquires all the previous elements
as ancestors, and the result is a {\it chain}, the causal set 
equivalent of
one-dimensional Minkowski space.  
At the other extreme, none of the elements has ancestors
(they are all ``spacelike'' to each other),
and the result is an {\it antichain},
a causal set which does not correspond to any spacetime (although it
can have an interpretation as analogous to a spacelike
hypersurface when it occurs embedded in a larger causal set.)  
In
between these extremes lie the more interesting regions in parameter
space, 
where one encounters,
for example,
CSG analogs of cyclical cosmologies, 
with coupling constants that get renormalized 
in such a way that the cosmos grows larger with each
successive cycle of collapse and re-expansion.

For present purposes,
the most important point is that the causal set
is analogous to a spacetime and the probabilities governing its growth
play the role of 
the
``law of motion'' for the spacetime (i.e. the Einstein equations in the
specific case of source-free gravity.)  
Of course, those of us working
with causal sets  hope that there's more to all of this than an analogy.  We
hypothesize that continuous spacetime is only an effective description
of a deeper reality,  a causal set whose
dynamics is described by something very like a CSG model.  
To be physically
realistic \dash and in particular to be able to generate a truly manifold-like
causal set  \dash this dynamics could not be classical; it would have to be
quantal in an appropriate sense.  A dynamical scheme of this sort is the
ultimate objective of current work, but even though we don't possess it
it yet, it is possible to imagine the kind of formalism to which it
would correspond mathematically, namely the formalism of ``generalized
quantum mechanics'' as codified in decoherence functionals and quantal
measure theory.   A dynamical scheme constructed along such lines
 would in the end be rather
similar to a CSG model.  The incorporation of interference (in the
quantum sense) would mark a dramatic difference, of course, but the
underlying kinematics or ``ontology'' would differ very little;
and even the
mathematical structure of the decoherence functional could find itself
in close analogy to the probability measure that defines a CSG
model.\footnote{$^\star$}
{Indeed, one 
 can obtain a non-classical
 decoherence functional by letting the
 parameters of the CSG model become complex.}
In particular the criterion of ``discrete general covariance''
 could carry over
essentially unchanged from the classical to the quantal case.  And since
considerations of 
Lorentz invariance and
general covariance seem to lie 
at the heart of the arguments
against a dynamical conception of reality, it seems fair enough to
reflect on them in the context of CSG models.  
Indeed, the cosmology of the CSG models is sufficiently realistic that
it's hard to imagine a question of principle relating to the
``being-becoming'' dichotomy that could not be posed in this
simplified context.

\noindent
3. To make my example more convincing
however, 
I
should probably try to explain a bit more in what sense a continuum
spacetime can emerge from a causal set.  Since this is essentially a
kinematical question, it can be answered fairly satisfactorily in the
present state of understanding.  Indeed, I claim that 
the correspondence between certain causal sets and certain spacetimes
is all I
really need to make my case, 
because once you accept it, 
all that remains is to
realize that a causal set can be generated by a process of 
``growth'' or
``birth'' in a way that does not presuppose any 
notion of distant simultaneity
or any concomitant notion of ``space developing in time''.

Perhaps a metaphor can bring out the key idea more clearly.  Think of
the causal set as an idealized growing tree 
(in the botanical sense,
not the combinatorial one).  Such a tree grows at the tips of its many
branches, and these sites of growth are independent of one another.
Perhaps a cluster of two leaves springs up at the tip of one branch (event
$A$) and at the same moment a single leaf unfolds itself at the tip of a
second branch (event $B$).  To a good approximation, the words ``at the
same moment'' make sense for real trees, but we know that they are not
strictly accurate, because events $A$ and $B$ occur at different
locations and distant simultaneity lacks objective meaning.  If the tree
were broad enough and the growth fast enough, we really could not say
whether event $A$ preceded or followed event $B$.  The same should be
true for the causal set.  It is ``growing at the tips'' but not in a
synchronized manner with respect to any external time.  There is no
single ``now'' that spreads itself over the entire process.\footnote{$^\dagger$}
{Mili{\v c} {\v C}apek [5] has proposed a musical metaphor for
 essentially the 
 same idea: that of a fugue.  In such a composition, each voice can seem
 for a while to unfold in its own region of space, its notes neither
 later nor earlier than the other's, until the musical lines come
 back together and intersect.}

``But wait a minute'', you might object.  ``Didn't you just describe the
CSG growth process as a succession of births in a definite order, and
doesn't the resulting ranking 
of the elements of $C$ imply something
akin to a distant simultaneity?''  
The
answer to this objection is that a definite birth-order,  or an ``external time'',
did figure in the description I gave, but it is to
be regarded as an artifact of the description analogous to one's choice
of coordinates for writing down the Schwarzschild metric.  Only insofar
as it reflects the intrinsic causal order of the causal set is this
auxiliary time objective.  The residue is ``pure gauge''.  Thus, any other
order of birth which is compatible with the intrinsic precedence
relation $\prec$ is to be regarded as physically equivalent to the
first, in the same sense that two diffeomorphic metrics are physically
equivalent.
So even though a CSG model rests on no background structure 
in the usual sense 
(unlike continuum gravity, 
where the underlying differentiable manifold acts as a background), 
one still meets with an issue very like that of
general covariance, 
stemming from the entry of an external time-parameter into the
mathematical definition of the model.  To complete my argument, then, I
will have to explain how this issue has been addressed
by the formalism,
but first let me
carry on with the task of explaining how a spacetime can emerge from a
causal set in the first place.

\subsection{Geometry =  number + order}
4. What might a stochastic process of the CSG type have to do with
spacetime and geometry, given that the type of mathematical object
involved (a past-finite poset), is not only discrete, but is at first
sight far removed from anything like a four-dimensional manifold?
Of course, the idea is that the continuity of
spacetime is illusory, that spacetime itself is only an emergent
reality, and that its inner basis is a causal set.
In order for this to be the case, the apparently rather primitive
structure of a discrete partial order must nonetheless conceal within
itself the type of information from which a Lorentzian geometry can be
recovered naturally, so that causal sets, or at least certain causal
sets can be placed in correspondence with certain spacetimes.
Fortunately the basis of this correspondence is easy to understand, at
least in broad outline.

With respect to a fixed system of coordinates, 
the spacetime 
 metric appears as a
symmetric matrix $g_{\alfa\beta}(x)$ of Lorentzian signature.  It is
therefore described by 10 real functions of the coordinates, 
$g_{00}(x)$, $g_{01}(x)$, \dots  $g_{33}(x)$.
Of these, the combination $g_{\alfa\beta}/|\det{g}|^{1/4}$
is determined if we know the light cones 
(i.e. the solutions of $g_{\alfa\beta}v^\alfa v^\beta=0$),
and the remaining factor of $\det(g)$ is determined if we know the {\it
volumes} of arbitrary spacetime regions $R$ since these are given by
integrals of the form $\int_R\sqrt{-\det(g)}d^4x$.
But we know the light cones once we know the causal ordering among the
point-events of spacetime, or in other words which point-events can
influence which others (the spacetime being assumed to carry a
time-orientation).  
Now let us postulate 
($i$) that this causal ordering directly reflects the ancestral relation
$\prec$ in the causal set; and
($ii$) that spacetime volume directly reflects the {\it number} of
causal set elements going to make up the region in question
(the number of births ``occurring in it'').  
We then have the ingredients
 for constructing 
 a four-geometry $M$, 
and if the construction succeeds, 
we may say that resulting $M$ is a good
approximation to the underlying causal set $C$: $M\approx C$.
When this is the case, $C$ may be identified with a subset of $M$, and
it turns out to be important 
(for questions like locality and Lorentz invariance) 
that this subset needs to be randomly distributed in
order to honor the postulate that number = volume. 

\subsection{Implications of growth models}
5. Having introduced the dynamics of sequential growth, and having pointed
out that a causal set growing in accord with such a model is capable in
principle of yielding a relativistic spacetime (not exactly, but to a
sufficient approximation), I am tempted to stop at this point and let
the example speak for itself.  On one hand sequential growth seems to me
to manifest ``becoming'' to the extent that any mathematical model can.
It even provides an objective correlate of our subjective perception of
``time passing''
in the unceasing cascade of birth-events that build up the causal set, 
by ``accretion'' as it were.\footnote{$^\flat$}
{The notion of ``accretive time'' that arises here seems close
 to that of C.D. Broad, and also to that of the ``Vibhajavadin'' school
 within the Buddhist philosophical tradition.}
On the other
hand, there is nothing in the model corresponding to a three-dimensional
space ``evolving in time''.  Rather, one meets with something which is
``four dimensional'' from the very beginning, but which at any stage of
its growth is still {\it incomplete}.\footnote{$^\star$}
{In this paper, I am using ``four-dimensional'' as a shorthand for 
 ``of a spacetime character, as opposed to a purely spatial character''.  
 There is nothing in the definition of a causal set that limits it to a
 dimensionality of four, or indeed to any uniform dimensionality.}
It's true that if we stop the process at any stage, we can identify the
maximal elements of $C$,
and these form a kind of ``future boundary'' of the growing causal set.  
But this ``boundary''
$A$ is an antichain, and as such can support intrinsically none of the
metrical structures of physical space.  It is only by reference to the many
relations of causal precedence connecting the elements of $A$ to their
ancestors that geometrical attributes can be attached to $A$ at all.\footnote{$^\dagger$}
{Arguably, a reference to the enveloping spacetime is present in the
 continuum as well, but there it is disguised by the fact that one need
 refer only to an arbitrarily small neighborhood of a hypersurface in
 order to
 define, for example, its extrinsic curvature, 
 whence reference to any {\it specific}
 earlier or later point can be deemed irrelevant.}
[6]  Thus, the spacetime character is primary in
CSG models, and any approximate notion of ``spacelike hypersurface'' is
derived from it.
Besides, stopping the process at a given stage has no objective meaning
within the theory, because with a different choice of 
birth-order,
the
causet at the same stage of growth would look entirely different.

The example of the CSG models seems to me to refute the contention that
relativistic spacetime is incompatible with genuine change, but I
suppose that no example can ever bring to a close
a debate that remains purely at the level
of interpretation.  If, on the other hand, we ask not whether becoming
is ``logically consistent'' with four-dimensionality, but rather whether
the combination of the two notions can be heuristically fruitful, then I
think that the development of the CSG models is in fact persuasive
evidence that it can.

The only other way to argue would be to refute,
one by one, 
the supposed proofs
that becoming and four-dimensionality exclude each other; but that would
have to be attempted by someone much better versed in those arguments
than I am.
Perhaps, however, it is fair to say that most 
such arguments
 presuppose that
the sole alternative to the ``block universe'' is a doctrine that
identifies reality with a three-dimensional instant.  
If this is so then the conception that emerges from the CSG models of a
four-dimensional, but still incomplete reality should be able, if not to
settle the debate, then at least to widen its terms in a fruitful
manner.

If we attend to our actual experience of time then no 
difficulty 
ever
arises, as pointed out long ago by Poincar{\'e}.  
Our ``now'' is
(approximately) localized and 
if we ask 
whether a distant event
spacelike to us
has or has not happened yet,
this question 
lacks intuitive sense.
But some ``opponents of becoming'' seem not to content themselves with
the experience of a ``situated observer''.
They want to imagine themselves as a ``super-observer'', 
who would take in all of existence at a glance.  
The supposition of such an observer {\it would} lead to a distinguished
``slicing'' of the causet, 
contradicting the principle
that such a slicing lacks objective meaning (``covariance'').  
Super-observers do not exist however, and the attempt to put ourselves
into their shoes brings 
the localized human experience of ``the now''
into 
conflict with the asynchronous multiplicity of ``nows''
of a CSG model
(cf. the analogy of the growing tree).  

\noindent
6. Returning from metaphor to mathematics, I would
like to deal briefly with two related features of
relativity-theory that arguments against `becoming' seem to rely upon
at a technical level, 
namely Lorentz invariance and general covariance.  
To what extent might the concept of a growing causet clash with
these features?
With respect to 
the first,
one can say something definite and quite
rigorous: one can quote a theorem.  With respect to the second, it is
less easy to reach a sure conclusion since 
the meaning of general
covariance in the context of a discrete and stochastic theory is still
more elusive than it is in a continuous and deterministic setting.

\subsection{Lorentz transformations}
In speaking of Lorentz invariance we remain essentially at the level of
kinematics, because,
so far as one knows,
the extant CSG models can give rise to significant
portions of Minkowski spacetime $\Minkowski^4$ only with vanishingly
small probability.\footnote{$^\flat$}
{However, there seem to exist more general Markov processes that do
 produce, for example, the future of the origin in $\Minkowski^4$
 [7].  These processes respect discrete general
 covariance in the sense of [1] but not Bell causality.}
Let us suppose, nevertheless, that some quantal growth process has
produced a causet $C$ resembling a large region of flat spacetime which
we can idealize as being all of Minkowski space: $C\approx\Minkowski^4$.
By definition, such a causal set is embeddable ``randomly'' in
$\Minkowski^4$, just as if it had been created by running a Poisson
process in $\Minkowski^4$.  With respect to the Poisson
probability-measure one can then prove that, with probability unity, it
is impossible to deduce a distinguished timelike direction from the
embedding [8].  In this sense, Lorentz invariance is
preserved exactly by the causal set, and one sees again how artificial
would be its decomposition into any sequence of antichains or other
analogs of space-at-a-moment-of-time.

\subsection{General covariance}
7. Finally, let us return to the question of general covariance.  In the
familiar context of continuum relativity, this phrase has
a double
significance.  In the first place it implies that only
diffeomorphism-invariant quantities possess physical meaning (where
the word ``quantities'' can be replaced, according to taste by ``events,
``questions'', ``predicates'', or ``properties'').\footnote{$^\star$}
{Underlying this limitation is the thought that spacetime points 
 \dash or in this case elements of the causal set \dash 
 possess no
 individuality beyond what they inherit from their relations to each
 other.
 According to John Stachel, they have ``quiddity'' without ``haecceity''.}
Given a spacetime
metric, it is thus meaningful to ask for the maximum area of a black
hole horizon but not for the value of the gravitational potential at
coordinate radius 17.  But aside from thus conditioning our definition of reality,
general covariance also demands in the second place that
a theory's 
equations of motion
(or its action-functional)
be diffeomorphism-invariant.  
Of course these two facets of general covariance are closely connected.  
Because
 general relativity
 is not a stochastic theory, 
it distinguishes rigidly between metrics that do and do not solve its field
equations.  Consequently,
the second facet of covariance flows directly from the
first as a consistency condition, because it would be senseless to
identify two metrics one of which was allowed by the equations of motion
and the other of which was forbidden; and conversely, the kinematical
identification must be made if one wishes the dynamics to be
deterministic.  Thus, the first or ``ontological'' facet of general
covariance tends to coalesce with its second or ``dynamical'' facet. 

In relation to causal sets, the context changes because one is dealing
with discrete structures and stochastic dynamics.  If one 
trys to rethink
 the
meaning of general covariance in this context, one can perhaps
distinguish now three relatively independent facets.
At the kinematic (or ontological) level,
essentially the same words apply as in general relativity: the causal
set elements ``{carry no inner identifiers}'', 
so that what has physical meaning is
only the 
isomorphism equivalence class of the given poset $C$.  
General covariance for causets can thus be interpreted as invariance under
relabeling, in analogy to the interpretation of general covariance as
coordinate-invariance in the continuum.

But because the
theory is stochastic 
this label-independence 
does not impose any obvious consistency condition on the
assignment of probabilities (or eventually quantal amplitudes)
to causets.  
It simply implies
that the only probabilities with physical meaning
are those attached to isomorphism equivalence classes of causets. 
In the CSG models, this is made precise by beginning with a
probability measure $\tilde\mu$ on a space $\tilde\Omega$ of 
{\it labeled} causets, 
and  passing from it to the induced measure $\mu$ on
the quotient space $\Omega$ whose members 
are the equivalence classes [3].
In effect the probability of an element of $\Omega$ is the sum of the
probabilities of all the members of that equivalence class.\footnote{$^\dagger$}
{In order to define $\mu$ consistently, one must take $\tilde\Omega$ to
 be a space of infinite causets, ones for which the growth process has
 ``run to completion''.  We meet here with an echo of the
 block-universe idea, that is in effect built into mathematicians'
 formalisation of the concept of stochastic process.}

The passage from $\tilde\Omega$ to $\Omega$ expresses (discrete) general
covariance kinematically and the induced passage from $\tilde\mu$ to
$\mu$ expresses it dynamically.  But what would correspond here to the
invariance of the equations of motion  \dash 
and do we require any such condition?  
In the analogous situation of gauge theories in $\Minkowski^4$, one
often ``fixes the gauge'', but then takes care to integrate the
resulting \dash non gauge-invariant \dash probability measure over entire
gauge equivalence classes.  (I am thinking of the
Fadeev-Popov approach to the Wick-rotated quantum field theory.)  This
would be the analog of letting $\tilde\mu$ depend on the labeling but
only computing 
probabilities 
 with respect to $\mu$.
However, before fixing the gauge one had a measure which {\it was}
gauge invariant (though defined only formally), 
and this invariance is a
crucial physical input to the theory.  
One might thus expect, in the causet case, that $\tilde\mu$ should itself
be relabeling invariant,  for only in this way would the classical
limit of the corresponding quantum theory have a chance to reproduce the
Einstein equations.  
It turns out that a natural invariance condition of this sort
can be found, and it is one of the two key inputs to the CSG models. 
The specific condition\footnote{$^\flat$}
{Notice that this condition of ``discrete general covariance'' is not
 itself formulated covariantly.  Notice also that it says something less
 than the following formal statement: ``the probability of a completed
 causet $C$ (a causet of countably many elements) is independent of its
 labeling.  This distinction was brought to light by Graham Brightwell,
 who also pointed out that this stronger statement {\it does} hold in
 the CSG models, even though it's not implied by discrete general
 covariance alone.
 One should also mention here
 an important difference between diffeomorphism-invariance
 and relabeling-invariance: the former is expressed by an invariance
 {\it group} arising as
 the automorphism group of a background structure (the manifold); the latter is not a
 group (a given permutation need not preserve naturality of the labeling)
 and there's no background (unless you count the integers 
 $\NaturalNumbers$
 from which our
 labels or parameter time $n$ come, but even if you do count them, their 
 automorphism group is trivial and does not generate relabelings).}
states that for any finite causet $C$ of
cardinality $n$, the probability to arrive at $C$ after $n$ births is
independent of the birth order of $C$'s elements (provided of course
that we limit ourselves to orderings that can actually happen,
i.e. to so called natural labelings of $C$).
Since, by construction, the CSG models fulfill this condition, one can
conclude that in these models there is no clash: 
discrete general covariance coexists
 harmoniously with the concept of a
dynamically growing causal set. 



%
\noindent
8. 
In the CSG models, a form of spatiotemporal discreteness plays a prominent
role, and for that reason alone,
one might question whether
the example with which we've
been working carries over 
to the spacetime continua of
special and general relativity.  
On the other hand,
the conception of a dynamically growing reality 
arguably
retains its meaning
in a continuous setting,
even if it
stretches one's intuition to a greater extent there.
Accordingly,
one might decide that ``becoming'' is not, 
after all, 
in conflict with
the four-dimensional Lorentzian manifold of
Relativity Theory.  
In that case,  
their compatibility would be something that
we might have recognized much earlier, without ever taking causal sets
into consideration.
The simplifying hypothesis of a discrete spatiotemporal substructure
would  
have served only as an inessential aid to our thinking.

On the other hand, one might in the end decide that a 
spacetime {\it continuum} necessarily is static, 
even though 
\dash as we have just seen \dash
a {\it  discrete} structure can consistently ``happen''.
In that case, 
an adherent of `becoming'
could claim that our intuition of time as a flow, 
had we but listened to it attentively, 
was all
along speaking to us of the discreteness of 
whatever process constitutes the inner basis of 
the phenomenon that we have been accustomed to conceptualizing as 
a spacetime continuum.


\bigskip
\noindent
Research at Perimeter Institute for Theoretical Physics is supported in
part by the Government of Canada through NSERC and by the Province of
Ontario through MRI.
This research was partly supported
by NSF grant PHY-0404646.

\ReferencesBegin                             

\ref [1] David P.~Rideout and Rafael D.~Sorkin,
``A Classical Sequential Growth Dynamics for Causal Sets'',
 \journaldata{Phys. Rev.~D}{61}{024002}{2000}
 \eprint{gr-qc/9904062}

\ref [2] 
                %
Madhavan Varadarajan and David Rideout,		
``A general solution for classical sequential growth dynamics of Causal Sets''
\journaldata {Phys. Rev. D} {73} {104021} {2006}
\eprint{gr-qc/0504066}

\ref [3] 
Graham Brightwell, Fay Dowker, Raquel S.~Garc{\'\i}a, Joe Henson and Rafael D.~Sorkin,
``{$\,$}`Observables' in Causal Set Cosmology'',
\journaldata{Phys. Rev.~D}{67}{084031}{2003} \lbr
\eprint{gr-qc/0210061}; 
\sepref
Graham Brightwell, {H. Fay Dowker}, {Raquel S. Garc{\'\i}a}, {Joe Henson} 
 and {Rafael D.~Sorkin},
``General Covariance and the `Problem of Time' in a Discrete Cosmology'',
 in K.G.~Bowden, Ed., 	
 {\it Correlations}, Proceedings of the ANPA 23 conference,
 held August 16-21, 2001, Cambridge, England 
 (Alternative Natural Philosophy Association, London, 2002), pp 1-17
\eprint{gr-qc/0202097}; 
\sepref
Fay Dowker and Sumati Surya,
``Observables in Extended Percolation Models of Causal Set Cosmology''
 \journaldata {Class. Quant. Grav.} {23} {1381-1390} {2006}
 \eprint{gr-qc/0504069}

\ref [4] 
Luca Bombelli, Joohan Lee, David Meyer and Rafael D.~Sorkin, 
``Spacetime as a Causal Set'', 
  \journaldata {Phys. Rev. Lett.}{59}{521-524}{1987}

\ref [5] Mili{\v c} {\v C}apek, 
{\it The philosophical impact of contemporary physics}
(Van Nostrand, 1961)

\ref [6] 
Seth Major, David Rideout, Sumati Surya,
``Spatial Hypersurfaces in Causal Set Cosmology''
 \journaldata {Class.Quant.Grav.} {23} {4743-4752} {2006}
 \eprint{gr-qc/0506133}; 
\sepref
Seth Major, David Rideout, and Sumati Surya
``On Recovering Continuum Topology from a Causal Set''
\eprint{gr-qc/0604124}

\ref [7] Graham Brightwell, unpublished notes on
general covariance.

\ref [8] 
Luca Bombelli, Joe Henson and Rafael D. Sorkin,
``Discreteness without symmetry breaking: a theorem''
(in preparation)
\eprint{gr-qc/0605006}

\end               


(prog1    'now-outlining
  (Outline 
     "\f......"
      "
      "
      "
   ;; "\\\\message"
   "\\\\Abstrac"
   "\\\\section"
   "\\\\subsectio"
   "\\\\appendi"
   "\\\\Referen"
   "\\\\ref....[^|]"
  ;"\\\\ref....."
   "\\\\end